\documentclass[prl,reprint,showpacs,floatfix]{revtex4-1}
\usepackage{graphicx,color,graphics,amsmath}

\begin{document}

\title{
Parity violating superfluidity in ultra-cold fermions \\
under the influence of artificial non-Abelian gauge fields
}

\author{
Kangjun Seo$^{1,2}$, Li Han$^1$ and C. A. R. S{\'a} de Melo$^1$
}

\affiliation{
1. School of Physics, Georgia Institute of Technology, Atlanta,
Georgia 30332, USA
}

\affiliation{
2. Department of Physics, Clemson University, Clemson,
South Carolina 29634, USA
}

\date{\today}

\begin{abstract}
We discuss the creation of parity violating Fermi superfluids in the
presence of non-Abelian gauge fields involving spin-orbit coupling
and crossed Zeeman fields. We focus on
spin-orbit coupling with equal Rashba and Dresselhaus
(ERD) strengths which has been realized experimentally in ultra-cold atoms,
but we also discuss the case of arbitrary mixing of Rashba and Dresselhaus
(RD) and of Rashba-only (RO) spin-orbit coupling. To illustrate the
emergence of parity violation in the superfluid, we analyze first
the excitation spectrum in the normal state and show that
the generalized helicity bands do not have inversion symmetry
in momentum space when crossed Zeeman fields are present.
This is also reflected in the superfluid phase, where the order
parameter tensor in the generalized helicity basis violates parity.
However, the pairing fields in singlet and triplet channels of
the generalized helicity basis are still parity even and odd,
respectively. Parity violation is further
reflected on ground state properties such as the spin-resolved momentum
distribution, and in excitation properties such as
the spin-dependent spectral function and density of states.
\end{abstract}


\maketitle

%
%

Parity violating phenomena are very rare in physics,
but a classical example is known from particle physics,
where parity violating processes of the weak interaction
were proposed~\cite{lee-yang-1956}
and observed in the decay of $^{60}$Co several decades
ago~\cite{wu-1957}.
In this case, the weak interactions allow for parity violation,
but the particle kinetic energies are parity even, reflecting the
inversion symmetry of their space. The Standard Model of particle
physics, which is a non-Abelian gauge theory, incorporates parity violations
and postulates that for nuclear beta decay parity is maximally violated.
Other examples of parity violation exist for instance in
condensed matter physics, where parity breaking is associated
with crystals without inversion symmetry~\cite{tilley-2006}
or with crystals which have inversion symmetry initially,
but can develop spontaneously permanent electric
polarization through lattice distortions leading to
ferroelectric materials~\cite{jona-shirane-1993}.
However, examples of parity breaking in superfluids, such as those encountered
in nuclear, atomic, condensed matter and astrophysics are hard to find,
and to our knowledge there seems to be no confirmed example in nature.

Recently, it has been possible to create non-Abelian gauge fields
in ultra-cold atoms via artificial spin-orbit (SO)
coupling of equal superposition of Rashba~\cite{rashba-1960}
$
{\bf h}_R ({\bf k})
=
v_R
(
- k_y {\hat {\bf x}}
+
k_x {\hat {\bf y}}
)
$
and Dresselhaus~\cite{dresselhaus-1955}
$
{\bf h}_D ({\bf k})
=
v_D
(
k_y {\hat {\bf x}}
+
k_x {\hat {\bf y}}
)
$
terms,
leading to the equal-Rashba-Dresselhaus (ERD)
form~\cite{spielman-2011, chapman-sademelo-2011}
${\bf h}_{ERD} ({\bf k}) = v {k_x} {\hat{\bf y}}$,
where $v_R = v_D = v/2$, for which parity preserving
superfluidity is possible~\cite{han-2011, seo-2012a, seo-2012b}.
Other forms of SO fields, such as the Rashba-only or
Dresselhaus-only cases, require additional lasers and create further
experimental difficulties~\cite{dalibard-2010},
while several theory groups have investigated the Rashba-only
case~\cite{shenoy-2011, chuanwei-2011, zhai-2011, hu-2011}
due to the connection to earlier condensed matter
literature~\cite{gorkov-2001, yip-2002, kane-2005}.

The current Zeeman-SO Hamiltonian created in the laboratory is
\begin{equation}
\label{eqn:zeeman-spin-orbit}
{\bf H}_{ZSO} ({\bf k})
=
- h_z \sigma_z
-
\left[
h_y +  h_{ERD} ({\bf k})
\right]
\sigma_y
\end{equation}
for an atom with center-of-mass momentum ${\bf k}$ and
spin basis $|\uparrow \rangle$,
$|\downarrow \rangle$.
The fields
$h_z = - \Omega_R/2$,  $h_y = -\delta/2$,
and $h_{ERD} ({\bf k}) = v k_x$
can be controlled independently.
Here, $\Omega_R$ is the Raman coupling and $\delta$ is the detuning,
which can be adjusted to explore phase diagrams
as achieved in $^{87}$Rb experiments~\cite{spielman-2011}, or to study
the high-temperature normal phases of Fermi atoms~\cite{wang-2012,cheuk-2012}.

In this letter, we show that ultra-cold Fermi superfluids in the presence
of non-Abelian gauge fields consisting of artificially created
spin-orbit and crossed Zeeman fields described in
Eq.~(\ref{eqn:zeeman-spin-orbit}) can produce a parity violating
superfluid state when interactions are included.
However, unlike the case of the Standard Model where
parity breaking is driven by the weak force,
in our case, parity breaking is driven by the effects of the non-Abelian
gauge field on the kinetic energy.
To illustrate the lack of parity in physical observables,
we analyze spectroscopic quantities such as the elementary excitation
spectrum, momentum distribution, spectral function and
density of states in the superfluid state.

%
%

{\it Hamiltonian:}
To analyze parity violation in ultra-cold Fermi superfluids,
we start from the Hamiltonian in momentum space as
\begin{equation}
\label{eqn:hamiltonian-0-momentum}
{\bf H}_0
=
\sum_{{\bf k} s}
\psi_s^\dagger ({\bf k})
{\bf H}_{0} ({\bf k})
\psi_s ({\bf k}),
\end{equation}
where
$
{\bf H}_{0} ({\bf k})
=
\left[
K ({\bf k}) {\bf 1}
-
{\bf h}_{\rm eff} ({\bf k})
\cdot
\mathbf{\sigma}
\right]
$
with $K({\bf k}) = {\bf k}^2/2m - \mu$ being
the single particle kinetic energy relative to the chemical potential
$\mu$; the vector-matrix $\mathbf{\sigma}$ describes the Pauli matrices
$(\sigma_x,\sigma_y, \sigma_z)$;
$
{\bf h}_{\rm eff} ({\bf k})
$
is the effective magnetic field with components
$
\left[
h_{x} ({\bf k}),
h_{y} ({\bf k}),
h_{z} ({\bf k})
\right]
$
and $\psi^\dagger_{s} ({\bf k})$ is the creation
operator for fermions with spin $s$ and momentum
${\bf k}$. In the ERD case, which is readily available in
ultra-cold atoms, the effective magnetic field is simply
$
{\bf h}_{\rm eff} ({\bf k})
=
\left[
0,
h_y +  h_{ERD} ({\bf k}),
h_z
\right],
$
where $h_y$ and $h_z$ are Zeeman components corresponding to the
detuning $\delta$ and the Raman coupling $\Omega_R$, while
$h_{ERD} ({\bf k})  = v k_x$ is the spin-orbit field.
We define the total number of fermions as
$N = N_\uparrow + N_\downarrow$,
and the induced population imbalance as
$P_{\rm ind} = (N_\uparrow - N_\downarrow)/N$.
We choose our scales through
the Fermi momentum $k_{F}$ defined from
$
N / V
=
k_{F}^3/(3\pi^2),
$
leading to the Fermi energy
$\epsilon_{F} = k_{F}^2/2m$
and the Fermi velocity $v_{F} = k_{F}/m$.

%
%
{\it Generalized Helicity Basis:}
The matrix ${\bf H}_0 ({\bf k})$ can be diagonalized
in the generalized helicity (GH) basis
$
\vert {\bf k}, \alpha \rangle
\equiv
\Phi_{\alpha}^\dagger ({\bf k}) \vert 0 \rangle
$
via a momentum-dependent SU(2) rotation generated
by the unitary matrix
\begin{equation}
\label{unitary-matrix}
{\bf U}_{\bf k}
=
\begin{pmatrix}
u_{\bf k} & v_{\bf k} \\
-v^*_{\bf k} & u_{\bf k}
\end{pmatrix},
\end{equation}
where the normalization condition
$
|u_{\bf k}|^2
+
|v_{\bf k}|^2
=
1
$
is imposed to satisfy the unitarity condition
$
{\bf U}_{\bf k}^\dagger
{\bf U}_{\bf k}
=
{\bf 1}.
$
The corresponding eigenvectors are the spinors
$
\Phi ({\bf k})
=
{\bf U}_{\bf k}^{\dagger}
\Psi ({\bf k}),
$
where
$
\Phi ( {\bf k} )
=
\left[
\Phi_\Uparrow ( {\bf k} ),
\Phi_\Downarrow ( {\bf k} )
\right]
$
is expressed in terms of
$
\psi ( {\bf k} )
=
\left[
\psi_\uparrow ( {\bf k} ),
\psi_\downarrow ( {\bf k} )
\right]
$
by the relations
$
\Phi_{\Uparrow}({\bf k})
=
u_{{\bf k}}  c_{{\bf k} \uparrow}
-
v_{{\bf k}}  c_{{\bf k} \downarrow}
$
and
$
\Phi_{\Downarrow}({\bf k})
=
v_{{\bf k}}^*  c_{{\bf k} \uparrow}
+
u_{{\bf k}}  c_{{\bf k} \downarrow}.
$
The coherence factor
$
u_{\bf k}
=
\sqrt{
\frac{1}{2}
\left(
1
+
\frac{h_{z}}
{\lvert
{\bf h}_{{\rm eff}}({\bf k})
\rvert}
\right)
}
$
is chosen to be real without
loss of generality, and
$
v_{\bf k}
=
- e^{i\varphi_{\bf k}}
\sqrt{
\frac{1}{2}
\left(
1 -
\frac{h_{z}}
{\lvert
{\bf h}_{\rm {eff}}({\bf k})
\rvert}
\right)
}
$
is a complex function with phase $\varphi_{\bf k}$
defined by
$
\varphi_{\bf k}
=
{\rm Arg}
\left[
h_\perp ({\bf k})
\right].
$
The complex field
$
h_\perp ({\bf k})
=
h_x ({\bf k}) - i h_y ({\bf k})
$
has components $h_x ({\bf k})$ and $h_y ({\bf k})$ along the $x$ and $y$
directions, respectively.
The magnitude of the effective field is
$
\vert {\bf h}_{\rm eff} ({\bf k}) \vert
=
\sqrt{
h_z^2
+
\vert h_\perp ({\bf k}) \vert^2
}.
$
In the ERD case
$h_x ({\bf k}) = 0$,
and the ratio
$
h_\perp ({\bf k})/\vert h_\perp ({\bf k}) \vert
=
e^{i \varphi_{\bf k}}
=
-i {\rm sgn} \left[ h_y ({\bf k}) \right],
$
where
$
h_y ({\bf k})
=
h_y
+
vk_x.
$

\begin{figure} [htb]
\includegraphics[width = 1.0\linewidth]{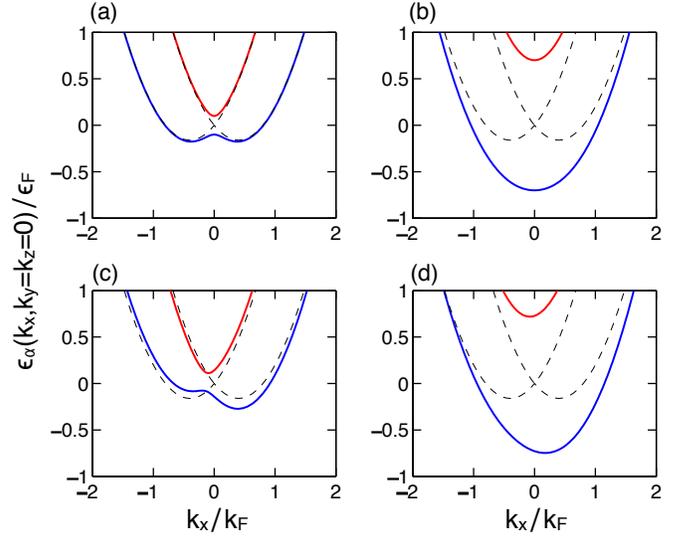}
\caption{
\label{fig:one}
(color online)
Generalized helicity bands $\epsilon_\Uparrow ({\bf k})/\epsilon_F$
(blue line) and $\epsilon_\Downarrow ({\bf k})/\epsilon_F$ (red line)
versus momentum $k_x/k_F$ with $k_y = k_z = 0$ and
for ERD spin-orbit coupling $v/v_F = 0.4$.
The black dashed lines show the helicity bands
for $v/v_F = 0.4$ with $h_z/\epsilon_F = h_y/\epsilon_F = 0$.
The Zeeman fields are
(a) $h_y/\epsilon_F = 0$ and $h_z/\epsilon_F = 0.1$,
(b) $h_y/\epsilon_F = 0$ and $h_z/\epsilon_F = 0.7$,
(c) $h_y/\epsilon_F = 0.1$ and $h_z/\epsilon_F = 0.1$,
(d) $h_y/\epsilon_F = 0.2$ and $h_z/\epsilon_F = 0.7$.
Notice that $\epsilon_{\alpha} ({\bf k}) \ne
\epsilon_{\alpha} (-{\bf k})$ in (c) and (d), indicating
the absence of parity.
}
\end{figure}

The generalized helicity spins $\alpha = (\Uparrow, \Downarrow)$
are aligned or antialigned with respect to the effective
magnetic field
$
{\bf h}_{\rm eff} ({\bf k}),
$
and the corresponding eigenvalues of
${\bf H}_0 ({\bf k})$
are
$
\xi_{\Uparrow} ({\bf k})
=
\epsilon_{\Uparrow} ({\bf k})  - \mu
$
and
$
\xi_{\Downarrow} ({\bf k})
=
\epsilon_{\Downarrow} ({\bf k}) - \mu.
$
Here, the  helicity energies are simply
$
\epsilon_{\Uparrow} ({\bf k})
=
K ({\bf k})
-
\vert h_{\rm eff} ({\bf k}) \vert
$
and
$
\epsilon_{\Downarrow} ({\bf k})
=
K ({\bf k})
+
\vert h_{\rm eff} ({\bf k}) \vert.
$
In the specific case of ERD coupling with non-zero
detuning $(h_y \ne 0)$ the effective field is
$
{\bf h}_{\rm eff} ({\bf k})
=
h_z {\hat {\bf z}}
+
\left[
h_y + h_{ERD} ({\bf k})
\right]
{ \hat {\bf y} },
$
with magnitude
$
\vert h_{\rm eff} ({\bf k}) \vert
=
\sqrt{
h_z^2
+
\left(h_y + v k_x\right)^2}
$
and parity violation occurring along the $x$ axis.
This is illustrated in Fig.~\ref{fig:one}, where
for finite $h_y$ (non-zero detuning $\delta$)
the generalized helicity bands
$
\epsilon_{\Uparrow} ({\bf k})
$
and
$
\epsilon_{\Downarrow} ({\bf k})
$
do not have well defined parity in momentum space.
As seen in Fig.~\ref{fig:one}(a)-(b),
parity is preserved for $v \ne  0$ if $h_y = 0$ (zero detuning).
While as noted in Fig.~\ref{fig:one}(c)-(d),
parity is violated for $v \ne  0$, if $h_y \ne 0$ (finite detuning).
Similar parity violation along the $x$ axis occurs for other mixtures
of Rashba and Dresselhaus terms as long as $h_y \ne 0$.

{\it Interactions and Order Parameter:}
In order to understand the underlying physics of this system, it
is important to rewrite the interaction Hamiltonian in the generalized helicity
basis. The starting interaction is
$
{\cal H}_I
=
-g
\sum_{\bf q}
b^\dagger ({\bf q})
b ({\bf q}),
$
where the pair creation operator with center
of mass momentum ${\bf q}$ is
$
b^\dagger ({\bf q})
=
\sum_{{\bf k}}
\psi^\dagger_{\uparrow} ( {\bf k} + {\bf q}/2 )
\psi^\dagger_{\downarrow} (-{\bf k} + {\bf q}/2 ),
$
can be written in the helicity basis as
$
{\widetilde {\cal H}}_{I}
=
-g
\sum_{{\bf q} \alpha \beta \gamma \delta}
B^\dagger_{\alpha \beta} ({\bf q})
B_{\gamma \delta} ({\bf q}),
$
where the indices $\alpha, \beta, \gamma, \delta$ cover
$\Uparrow$ and $\Downarrow$ states. Pairing is now described
by the operator
\begin{equation}
\label{eqn:pairing-operator-helicity-basis}
B_{\alpha \beta} ({\bf q})
=
\sum_{\bf k}
\Lambda_{\alpha \beta}
({\bf k}_+, {\bf k}_-)
\Phi_{\alpha} ({\bf k}_+ )
\Phi_{\beta}  ({\bf k}_-)
\end{equation}
and its Hermitian conjugate, with momentum indices
${\bf k}_\pm = \pm {\bf k} + {\bf q}/2$.
The matrix
$
\Lambda_{\alpha \beta}
({\bf k}_+, {\bf k}_-)
$
is directly related to products of coherence factors
$u ({\bf k}_{\pm} )$, $v ({\bf k}_{\pm} )$ (and their
complex conjugates) of the momentum
dependent SU(2) rotation matrix ${\bf U} ({\bf k}_{\pm}) $.
Seen in the GH basis,
the interactions reveal that the center of mass momentum
${\bf k}_{+} + {\bf k}_{-} = {\bf q}$ and
the relative momentum ${\bf k}_{+} - {\bf k}_{-} = 2{\bf k}$
are coupled and no longer independent, and thus do not
obey Galilean invariance. The interaction constant $g$ is
related to the scattering length via the Lippman-Schwinger relation
$V/g = -Vm/(4\pi a_s) + \sum_{\bf k}1/(2\epsilon_{\bf k}).$

From Eq.~(\ref{eqn:pairing-operator-helicity-basis}) it is
clear that pairing between fermions of momenta
${\bf k}_{+}$ and ${\bf k}_{-}$ can occur within the
same helicity band (intra-helicity pairing)
or between two different helicity bands
(inter-helicity pairing). For pairing at zero center-of-mass
momentum ${\bf q} = 0$,
the order parameter for superfluidity is the tensor
$
\Delta_{\alpha \beta} ({\bf k})
=
\Delta_0
\Lambda_{\alpha \beta} ({\bf k}, -{\bf k}),
$
where
$
\Delta_0
=
- g
\sum_{\gamma \delta}
\langle
B_{\gamma \delta} ({\bf 0})
\rangle,
$
leading to
components:
$
\Delta_{\Uparrow \Uparrow} ({\bf k})
=
\Delta_0
\left(
u_{\bf k} v_{-\bf k}
-
v_{\bf k} u_{-\bf k}
\right)
$
for total helicity projection
$\lambda = +1$;
$
\Delta_{\Uparrow \Downarrow} ({\bf k})
=
-
\Delta_0
\left(
u_{\bf k} u_{-\bf k}
+
v_{\bf k} v_{-\bf k}^*
\right)
$
and
$
\Delta_{\Downarrow \Uparrow} ({\bf k})
=
\Delta_0
\left(
u_{\bf k} u_{-\bf k}
+
v_{\bf k}^* v_{-\bf k}
\right)
$
for total helicity projection $\lambda = 0$; and
$
\Delta_{\Downarrow \Downarrow} ({\bf k})
=
\Delta_0
\left(
u_{\bf k} v_{-\bf k}^*
-
v_{\bf k}^* u_{-\bf k}
\right)
$
for total helicity projection $\lambda = -1$.
Parity is violated in $\Delta_{\alpha \beta} ({\bf k})$
since they do not have well defined parity for non-zero spin-orbit
coupling and crossed Zeeman fields $h_y$ and $h_z$.

However, we may still define singlet and triplet sectors
in the generalized helicity basis, which are even and odd
in momentum space respectively for any value of $h_y$.
The singlet sector is defined by the scalar order parameter
$
\Delta_{S, 0} ({\bf k})
=
\left[
\Delta_{\Uparrow \Downarrow} ({\bf k})
-
\Delta_{\Downarrow \Uparrow} ({\bf k})
\right]/2
$
corresponding to $\lambda = 0$.
While the triplet sector is defined by the vector
order parameter
$
\Delta_{T, \lambda} ({\bf k}),
$
by its generalized helicity components
$
\Delta_{T, +1} ({\bf k})
=
\Delta_{\Uparrow \Uparrow} ({\bf k})
$
corresponding to $\lambda = +1$;
$
\Delta_{T, 0} ({\bf k})
=
\left[
\Delta_{\Uparrow \Downarrow} ({\bf k})
+
\Delta_{\Downarrow \Uparrow} ({\bf k})
\right]/2
$
corresponding to $\lambda = 0$;
$
\Delta_{T, -1} ({\bf k})
=
\Delta_{\Uparrow \Uparrow} ({\bf k})
$
corresponding to $\lambda = -1$.

%
%

{\it Superfluid Ground State and Elementary Excitations:}
The ground state for uniform superfluidity
can be expressed in terms of fermion pairs
in the GH basis as the many-body wavefunction
$
\vert G \rangle
=
\prod_{\bf k}
\left\{
\sum_{\alpha \beta}
\left[
U_{\alpha \beta} ( {\bf k} )
+
V_{\alpha \beta} ( {\bf k} )
\Phi_{\alpha}^\dagger ( {\bf k} )
\Phi_{\beta}^\dagger ( - {\bf k} )
\right]
\right\}
\vert 0 \rangle,
$
where $\vert 0 \rangle$ is the vacuum state with no particles.

The Hamiltonian matrix in the GH basis is
\begin{equation}
\label{eqn:mean-field-hamiltonian-matrix}
{\widetilde {\bf H}}_{\rm ex} ({\bf k})
=
\begin{pmatrix}
\xi_{{\bf k}\Uparrow} & 0 &
\Delta_{\Uparrow\Uparrow}({\bf k}) & \Delta_{\Uparrow\Downarrow}({\bf k}) \\
0 & \xi_{{\bf k} \Downarrow}  &
\Delta_{\Downarrow \Uparrow}({\bf k}) & \Delta_{\Downarrow\Downarrow}({\bf k}) \\
\Delta_{\Uparrow\Uparrow}^* ({\bf k}) & \Delta_{\Downarrow\Uparrow}^*({\bf k})
& -\xi_{-{\bf k} \Uparrow} & 0 \\
\Delta_{\Uparrow\Downarrow}^*({\bf k}) & \Delta_{\Downarrow\Downarrow}^*({\bf k})
& 0 & -\xi_{-{\bf k}\Downarrow}
\end{pmatrix},
\end{equation}
which is traceless, showing that
the sum of its eigenvalues is zero.
We have obtained analytical solutions for the eigenvalues of
$\widetilde {\bf H}_{\rm ex} ({\bf k})$
for arbitrary RD spin-orbit orbit and arbitrary Zeeman fields $h_y$ and $h_z$,
but we do not list them here, because their expressions are cumbersome.
However, for each momentum ${\bf k}$, the determinant
$
{\rm Det}
\left[
\omega {\bf 1} -  \widetilde{\bf H}_{\rm ex} ({\bf k})
\right],
$
leads to the quartic equation
\begin{equation}
\label{eqn:eigenvalue-equation}
\omega^4
+
a_3 ({\bf k}) \omega^3
+
a_2 ({\bf k}) \omega^2
+
a_1 ({\bf k}) \omega
+
a_0 ({\bf k})
=
0.
\end{equation}

In the particular case of ERD spin-orbit coupling with crossed Zeeman
fields, the coefficients become $a_3 ({\bf k}) = 0$,
the coefficient of the quadratic term takes the form
$$
a_2 ({\bf k})
=
- 2
\left(
K^2({\bf k})
+ |\Delta_0|^2 + |v k_x|^2 + |h_y|^2 + |h_z|^2
\right),
$$
while the coefficient of the linear term is
$
a_1 ({\bf k})
=
-8  K({\bf k}) ( v k_x ) h_y,
$
and lastly the coefficient of the zero-th order term is
$$
a_0 ({\bf k})
=
\xi_{\Uparrow}({\bf k})
\xi_{\Downarrow}({\bf k})
\xi_{\Uparrow}({-\bf k})
\xi_{\Downarrow}(-{\bf k})
+
|\Delta_0|^2 \alpha_0^2 ({\bf k}),
$$
where
$
\alpha_0^2 ({\bf k})
=
\left(
 2 K^2({\bf k})
+ |\Delta_0|^2 + h_0^2 ({\bf k})
\right)
$
with
$
h_0^2 ({\bf k})
=
2|v k_x|^2 -2 |h_y|^2 -2 |h_z|^2.
$
Notice that $a_2 ({\bf k})$ and $a_0 ({\bf k})$
have even parity, while $a_1 ({\bf k})$ has odd parity and is thus
responsible for the parity violation that occurs in the elementary
excitation spectrum. Furthermore, parity violation occurs only
when both $v$ and $h_y$ are non-zero, since when either $h_y = 0$ or $v = 0$
the coefficient $a_1 ({\bf k})$ vanishes and parity in the elementary
excitation spectrum is fully restored. From the secular equation,
it follows that when $k_x = 0$, the coefficient
$a_1 ({\bf k})$ also vanishes and the excitation energies
$E_i (0, k_y, k_z)$ have the same analytical form as in the case for
$h_y = 0$, with the simple replacement of $h_z^2 \to h_z^2 + h_y^2$.
This property is just a consequence of the reflection symmetry of the
Hamiltonian through the $k_x = 0$ plane. However, parity is violated,
because inversion symmetry through the origin of momenta does not
exist, that is, $E_i (-{\bf k}) \ne E_i ({\bf k})$.
In contrast, quasiparticle-quasihole symmetry
is preserved since the corresponding quasiparticle-quasihole energies
obey the relations
$E_2 ({\bf k}) = -E_3 (-{\bf k})$ and $E_1 ({\bf k}) = -E_4 (-{\bf k})$.

A simple inspection shows that gapless and fully gapped phases emerge.
A gapless phase with two rings of nodes (US-2) appears
when
$
h_y^2 + h_z^2 - \vert \Delta_0
\vert^2
>
0
$
and
$\mu > \sqrt{h_y^2 + h_z^2 - \vert \Delta_0 \vert^2}$.
A gapless phase with one ring of nodes (US-1) occurs for
$h_y^2 + h_z^2 - \vert \Delta_0 \vert^2 > 0$ and
$
\vert \mu \vert
<
\sqrt{h_y^2 + h_z^2 - \vert \Delta_0 \vert^2}.
$
A directly gapped phase (d-US-0) arises for
$
h_y^2 + h_z^2 - \vert \Delta_0 \vert^2
>
0
$
and
$
\mu
<
- \sqrt{h_y^2 + h_z^2 - \vert \Delta_0 \vert^2},
$
while an indirectly gapped phase (i-US-0) emerges for
$
h_y^2 + h_z^2 - \vert \Delta_0 \vert^2
<
0
$
and $\mu > 0$.
Lastly, the quasiparticle excitation energy $E_2 ({\bf k})$ becomes negative
in certain momentum regions when $h_y^2 > \vert \Delta_0 \vert^2$, indicating
that the uniform ground state becomes less energetically favorable
against the normal state~\cite{footnote-1}.

%
%

\begin{figure} [htb]
\centering
\includegraphics[width=1.0 \linewidth]{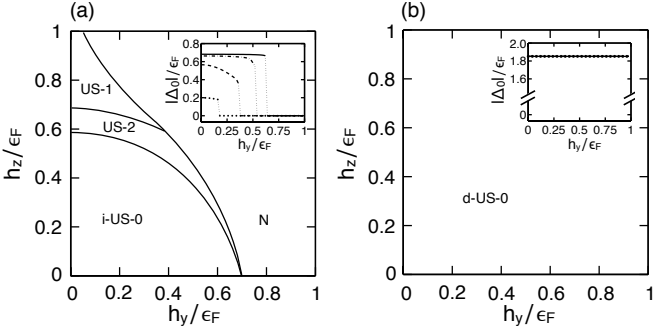}
\caption{
\label{fig:two}
The $T = 0$ phase diagram in the $h_y$-$h_z$ parameter
space showing various uniform superfluid phases US-2, US-1, d-US-0
and i-US-0, and the normal phase for ERD spin-orbit coupling
$v/v_F = 0.4$ and at (a) unitarity $1/(k_{F}a_s) = 0.0$ and in
(b) the BEC regime $1/(k_{F} a_s) = 2.0$.
The insets show $\vert \Delta_0 \vert$ as a function of $h_y$
for $h_z/\epsilon_F = 0.2$ (dotted line);
for $h_z/\epsilon_F = 0.4$ (dot-dashed line);
$h_z/\epsilon_F = 0.6$ (dashed line);
and $h_z/\epsilon_F = 0.8$ (solid line).
In the range shown, $\vert \Delta_0 \vert$ is essentially
independent of $h_y$ and $h_z$ in the BEC regime.
}
\end{figure}

{\it Phase Diagram and Thermodynamic Potential:}
From the thermodynamic potential
$
\Omega_{\rm US}
=
-(T/2) \sum_{ {\bf k},j }
\ln
\left[
1 +
\exp
\left(
-E_j ({\bf k})/T
\right)
\right]
+
\sum_{\bf k} K ({\bf k})
+
\vert \Delta_0 \vert^2/g
$
we obtain self-consistently the zero temperature $(T = 0)$
phase diagram as a function of crossed Zeeman fields $h_y$ and $h_z$
for $v/v_F = 0.4$ at unitarity $1/(k_F a_s) = 0$ in Fig.~\ref{fig:two}(a),
and at the BEC regime $1/(k_F a_s) = 2.0$ in Fig.~\ref{fig:two}(b),
but a stability analysis against non-uniform phases is necessary
as in the parity-preserving case~\cite{seo-2012a,seo-2012b}.
At unitarity the uniform superfluid phases i-US-0, US-1, US-2
and the normal (N) phase are present in the range shown, while
in the BEC regime only the d-US-0 occurs in the same range of fields.
The transitions between different US phases is topological
with no change in symmetry as in the parity-preserving
case~\cite{seo-2012a,seo-2012b}.
While the transitions from US phases
to the N phase involve a change in symmetry,
from broken to non-broken U(1), and are discontinuous,
as seen in the insets of Fig.~\ref{fig:two}.

%

%
%

\begin{figure} [htb]
\centering
\includegraphics[width=1.0 \linewidth]{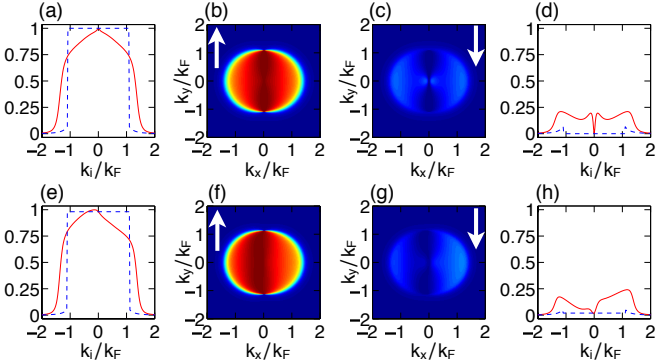}
\caption{ \label{fig:three}
(color online) Momentum distributions $(T = 0)$
$n_\uparrow ({\bf k})$  (two left-most columns) and
$n_\downarrow ({\bf k})$ (two right-most columns)
for $1/(k_F a_s) = 0.0$ and $v/v_{F} = 0.40$
at the US-1 phase. In (a)-(d) the field values
are $h_y/\epsilon_F = 0$, $h_z/\epsilon_F = 0.7$,
with $\mu/\epsilon_F = 0.5803$, $\vert \Delta_0 \vert/\epsilon_F = 0.3592$,
and $P_{\rm ind} = 0.6592$.
In (e)-(h) the field values are
$h_y/\epsilon_F = 0.2$ and $h_z/\epsilon_F = 0.7$,
with $\mu/\epsilon_F = 0.5871$, $\vert \Delta_0 \vert/\epsilon_F = 0.3157$,
and $P_{\rm ind} = 0.6958$.
The blue-dashed and red-solid lines represent cuts of
$ n_s ({\bf k}) $ along the directions
$(0, k_y, 0)$ and $(k_x, 0, 0)$, respectively.
}
\end{figure}

{\it Detecting parity violation:}
A direct measurement of parity violation in the superfluid
state can be made through the momentum distributions
$
n_s ({\bf k})
=
\langle
\psi^\dagger_s ({\bf k})
\psi_s ({\bf k})
\rangle.
$
They are illustrated in Fig.~\ref{fig:three} for US-1 superfluid
with spin-orbit $v/v_F = 0.4$ and interaction $1/(k_F a_s) = 0$,
in the parity-preserving case with
$h_y/\epsilon_F = 0$ and $h_z/\epsilon_F = 0.7$ in (a)-(d)
and  in the parity-violating case
with $h_y/\epsilon_F = 0.2$ and $h_z/\epsilon_F = 0.7$ in (e)-(h).
At finite temperatures, the momentum distributions broaden, but
parity violation is still self-evident.

\begin{figure} [htb]
\centering
\includegraphics[width=1.0 \linewidth]{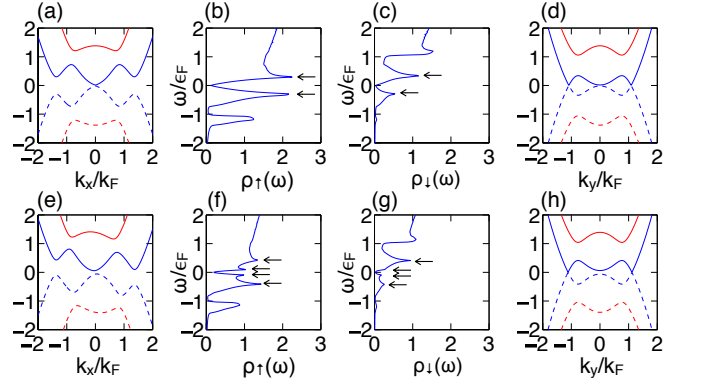}
\caption{
\label{fig:four}
(color online) Eigenvalues $E_i ({\bf k})$ and density of states $\rho_s (\omega)$
(in units of $\epsilon_F$ and $\epsilon_F^{-1}$, respectively) for $1/(k_F a_s) = 0$ and $v/v_{F} = 0.4$
in the US-1 phase, but close to the US-1/US-2 boundary,
with parameters
$h_y/\epsilon_{F} = 0$, $h_z/\epsilon_F = 0.7$,
$\mu/\epsilon_F = 0.5803$, $\vert \Delta_0 \vert/\epsilon_F = 0.3592$, and
$P_{\rm ind} = 0.6592$ in (a)-(d);
and
$h_y/\epsilon_{F} = 0.2$, $h_z/\epsilon_F = 0.7$,
$\mu/\epsilon_F = 0.5871$, $\vert \Delta_0 \vert/\epsilon_F = 0.3157$, and
$P_{\rm ind} = 0.6958$ in (e)-(h).
Cuts of $E_i ({\bf k})$ along $(k_x, 0, 0)$ are shown in (a) and (e),
and along $(0, k_y, 0)$ are shown in (d) and (h).
In panels for $\rho_{s} (\omega)$ a small broadening
$\delta/\epsilon_F = 0.01$ is used.  The black arrows indicate
examples of peaks that split when parity breaking occurs for
finite $h_y$.
}
\end{figure}

Parity violation is also manifested in other momentum resolved
properties such as the spectral function
$
{\cal A}_s (\omega, {\bf k})
=
-(1/\pi) {\rm Im} G_{ss} (i\omega = \omega + i\delta, {\bf k}),
$
where
$
G_{ss} (i\omega, {\bf k})
=
\left[
i\omega {\bf 1} - {\widetilde H}_{\rm ex} ({\bf k})
\right]^{-1},
$
written in the $s = \uparrow, \downarrow$ basis.
Instead, in Fig.~\ref{fig:four}, we choose to illustrate  a manifestation
of parity violation in the elementary excitation spectrum for the
US-1 superfluid phase, and the corresponding implications for momentum
integrated quantities such as the spin-resolved density of states
$
\rho_s (\omega)
=
\sum_{\bf k}
{\cal A}_s (\omega, {\bf k}).
$
The most important point is that for finite spin-orbit coupling $v$
and when $h_y \ne 0$, the excitation energies
$E_i ({\bf k}) \ne E_i (-{\bf k})$. This implies that
degenerate peaks at $h_y = 0$ (corresponding to minima or maxima of
the excitation spectrum) are increasingly split with growing $h_y$.
This effect is illustrated in Fig.~\ref{fig:four}
at the locations indicated by the small black arrows.

%

%
%

{\it Conclusions:}
We showed that non-Abelian gauge fields consisting of spin-orbit and
crossed Zeeman fields lead to parity  violating superfluidity
in ultra-cold Fermi systems. We derived general relations that
can be applied to spin-orbit couplings
involving any linear combination of Rashba and Dresselhaus terms.
We focused mostly on the case of equal Rashba-Dresselhaus (ERD) spin-orbit
coupling. The presence of such fields produce a superfluid order
parameter tensor whose components in the generalized helicity basis
are neither even nor odd under spatial inversion. Even though the elements of
this tensor written in generalized singlet or triplet helicity channels
have even or odd parity, respectively, the excitation
spectrum does not have well defined parity, but preserves
quasiparticle-quasihole symmetry.
This parity violation has important experimental
signatures leading to momentum distributions
without inversion symmetry and
to spin-resolved density of states that possess split peaks
in frequency.

\acknowledgements{We thank ARO (W911NF-09-1-0220) for support.}

\end{document}